\documentclass[a4paper]{article}

\usepackage{icrc2013}
 \usepackage[english]{babel}

\title{Estimation of the TeV gamma-ray duty cycle of Mrk 421 with Milagro}
\shorttitle{Estimation of the TeV $\gamma$-ray duty cycle of Mrk 421 with Milagro}

\authors{
B. Patricelli$^{1}$,
M.M. Gonz\'alez$^{1}$,
N. Fraija$^{1}$,
A. Marinelli$^{2}$
for the Milagro Collaboration.
}

\afiliations{
$^1$ Instituto de Astronom\'ia, UNAM, M\'exico D.F., 04510, M\'exico \\
$^2$ Instituto de Fisica, UNAM, M\'exico D.F., 04510, M\'exico \\
}

\email{bpatricelli@astro.unam.mx}

\abstract{The blazar Markarian 421 (Mrk 421) is one of the brightest sources in the extragalactic X-ray/TeV sky.  It is also one of the fastest varying TeV $\gamma$-ray sources, showing flaring activity on time scales as short as tens of minutes. To know the level of activity of this source, Tluczykont et al. (2007) \cite{2007JPhCS..60..318T} calculated the fraction of time spent by Mrk 421 in flaring states with fluxes above 1 Crab at TeV energies (i.e., TeV - duty cycle). Here we present an alternative approach to calculate the TeV duty cycle of Mrk 421 taking advantage of the continuous monitoring of the source by the Milagro observatory. Milagro was a water Cherenkov detector sensitive at energies between 100 GeV and 100 TeV. We present our estimation of the TeV - duty cycle and  study its robustness.}

\keywords{VHE gamma-rays, blazars, duty cycle.}

\begin{document}
\maketitle

\section{Introduction}\label{sec:intro}
 Blazars are a subclass of active galactic nuclei (AGN) characterized by broadband non-thermal emission from radio to very high energies (VHE, E $>$ 100 GeV, Horns 2008 \cite{2008RvMA...20..167H}). They show strong flux variability at almost all frequencies of the spectral energy distribution on different time scales, from minutes (see, e.g., Aharonian et al. 2007 \cite{2007ApJ...664L..71A}) to months (see, e.g., von Montigny et al. 1995 \cite{1995ApJ...440..525V}). This large spread in time variability makes it difficult to quantify important parameters such as the duty cycle (DC). The DC is defined as the fraction of time spent in a high (``flaring'') state, thus, 
\begin{equation}\label{eq:DC}
DC=\frac{\sum_i t_i}{\sum_i t_i+T_{\rm baseline}}=\frac{T_{\rm flare}}{T_{\rm flare}+T_{\rm baseline}},
\end{equation}
where $t_i$ is the time that the source spends in a $i$ flaring state  and $T_{\rm baseline}$ is the total time in which the source is in the baseline flux state. The definition of a flaring state varies from author to author (see e.g. Krawczynski et al. 2004 \cite{2004ApJ...601..151K}, Wagner 2008 \cite{2008MNRAS.385..119W}). The baseline flux may be stable and constant with time, although it may present intrinsic variations. In the former case, a flaring state can be defined as any state with flux higher than the baseline flux. In the latter case, a flaring state must be defined taking into account the assumed or measured intrinsic variations of the baseline flux. The identification of a baseline level is also needed to identify the blazar quiescent level: without a proper baseline level, only an upper limit of the flaring flux can be determined (Wagner 2011 \cite{2011ICRC....8..147W}).

Mrk 421 is one of the closest (redshift z=0.03; de Vaucouleurs et al. 1991 \cite{1991rc3..book.....D}) and brightest blazars known; it was also the first BL Lac source detected at energies above 100 MeV by EGRET in 1991 (Lin et al. 1992 \cite{1992ApJ...401L..61L}) and the first extragalactic object detected in the TeV energy band, by the Whipple Collaboration (Punch et al. 1992 \cite{1992Natur.358..477P}). 

The duty cycle at TeV energies has been estimated for Mrk 421 by Tluczykont et al. (2007) \cite{2007JPhCS..60..318T}. They collected data from different imaging atmospheric Cherenkov telescopes (IACTs: HEGRA, HESS, MAGIC, CAT, Whipple and VERITAS) from 1992 to 2009. They combined the light curves from the different experiments converting the measured integral flux to flux values in units of the Crab Nebula flux and normalizing to a common energy threshold of 1 TeV and obtained a distribution of flux states for Mrk 421. Finally, they estimated the TeV duty cycle as the time that the source spent in a flaring state to the total observation time of the telescopes. They considered different flare flux thresholds. For a flare flux threshold of 1 Crab, they found a TeV $DC$ of $\sim$ 40 \%. This value may overestimate the true $DC$ since IACT observations are biased towards high flux states due to their external and self triggering on high states (Tluczykont et al. 2007 \cite{2007JPhCS..60..318T}). In the present paper, we use  the definition of flaring state as defined by Tluczykont, et al. (2007) \cite{2007JPhCS..60..318T} for a flux threshold of 1 Crab. This is a conservative definition and allows  us to compare our results with existing results \cite{2007JPhCS..60..318T}. We present a different approach with respect to Tluczykont et al. 2007 \cite{2007JPhCS..60..318T} to calculate the TeV $DC$ of Mrk 421, that take advantage of the continuous and unbiased long term monitoring of the source by the  Milagro detector.

\section{The Milagro detector}
Milagro (see Atkins et al. 2004 \cite{2004ApJ...608..680A}) was a large water-Cherenkov detector located in the Jemez Mountains near Los Alamos, New Mexico, USA at an altitude of 2630 m above sea level. It was designed to detect very high energy (VHE) gamma rays at energies between 100 GeV and 100 TeV (Abdo et al. 2008a,b \cite{2008ApJ...688.1078A,2008PhRvL.101v1101A}). It had a $\sim$2 sr field  of view and  a $\geq$ 90$\%$ duty cycle that allowed it to continuously monitor the entire overhead sky. It operated from 2000 to 2008. It was composed of a central 80 m $\times$ 60 m $\times$ 8 m water reservoir instrumented with 723 photomultiplier tubes (PMTs) arranged  in two layers. The top ``air-shower" layer consisted of 450 PMTs under 1.4m of purified water, while the bottom ``muon" layer had 273 PMTs located 6m below the surface. The air-shower layer was used to reconstruct the direction of the air shower by measuring the relative arrival times of the shower particles across the array. The muon layer was used to discriminate between gamma-ray induced and  hadron-induced air showers. In 2004, a sparse 200 m x 200 m array of 175 ``outrigger'' was added around the central reservoir. This array increased the area of the detector and improved the gamma/hadron separation. The detector reached its final configuration in September 2005. 

\section{TeV duty cycle of Mrk 421}
Milagro detected Mrk 421 in the period from September 21, 2005 to March 15, 2008 with a statistical significance of 7.1 standard deviations at a median energy of 1.7 TeV. From the analysis of the light curve we found (Abdo et al. 2013 \cite{MilagroMrk421}) that the Mrk 421 flux is consistent with being constant along the whole 3-year observation period, with an average value above 1 TeV of $\bar f$= ($2.05 \,\pm 0.30$) $\times 10^{-11}\,\rm{cm^{-2}\,s^{-1}}$ ($\chi^2$=134 for 122 degrees of freedom) equivalent to 0.85$\pm$0.13 Crab. This average flux results from combining the baseline flux,  $F_{\rm baseline}$, and the contributions of the fluxes of any other higher (``flaring'') state $i$, $f_{{\rm flare},i}$. Thus,
\begin{equation}\label{eq:fluence}
\bar{f} \times T_{\rm Milagro}= F_{\rm baseline}\times T_{\rm baseline}+\cal{F_{\rm flare}},
\end{equation}

where $T_{\rm Milagro}$ is the total observation period of Milagro given by $T_{\rm baseline}+T_{\rm flare}$  and $\cal{F_{\rm flare}}$ is the total fluence of all high states given by $\sum_i f_{{\rm flare},i}\, t_i$ 

The knowledge of $\bar f$ is not enough to estimate the TeV $DC$, as the same value of $\cal{F_{\rm flare}}$ could be obtained by considering many long-duration low-flux flares or a few short-duration high-flux flares, leading to different $DC$ values. Therefore, a distribution of flux states of Mrk 421 is needed. We used the same distribution of flux states above 1 TeV used by Tluczykont et al. 2007 \cite{2007JPhCS..60..318T}. This distribution is well fit by a function $f(x)$\footnote{The variable $x$ represents the flux of Mrk 421 above 1 TeV in Crab unit.} which is the sum of  a Gaussian component, describing the baseline flux state plus a {\rm log}-normal function, describing the flaring states (Tluczykont et al. 2010 \cite{2010A&A...524A..48T}). The mean of the Gaussian component ($\sim$ 0.33 Crab) represents an upper limit on the value of $F_{\rm baseline}$ (Tluczykont et al. 2010 \cite{2010A&A...524A..48T}) because some lower fluxes may be missing. The detectors may not be sensitive enough to detect them for short observation periods.

We used the distribution of Tluczykont et al. 2010 \cite{2010A&A...524A..48T} to calculate the average flare flux, \mbox{$<f_{\rm flare}>$}, given by,

\begin{equation}\label{eq:ffm}
<f_{\rm flare}>=\frac{\int_{1 \, \rm{Crab}}^{F_{\rm lim}} x\,f(x)\,dx}{\int_{1 \, \rm{Crab}}^{F_{\rm lim}} f(x)\,dx}
\end{equation}

with $F_{\rm lim}$ the maximum flux considered in the distribution, i.e. $F_{\rm lim}$=10 Crab \cite{2010A&A...524A..48T}.  Then, we have $<f_{\rm flare}>$= 2.64 Crab.

${\cal F_{\rm flare}}$ can be written in terms of $<f_{\rm flare}>$ as following,

\begin{equation}\label{eq:fluencehigh}
{\cal F_{\rm flare}}=<f_{\rm flare}>\times T_{\rm flare}.
 \end{equation}
By inserting Eq. \ref{eq:fluencehigh} in Eq. \ref{eq:fluence} we obtain
\begin{equation}\label{eq:tflare}
T_{\rm flare}=\frac{\left( \bar{f}-F_{\rm baseline}\right) T_{\rm Milagro}}{<f_{\rm flare}>-F_{\rm baseline}}
\end{equation}
Then, Eq. \ref{eq:DC} becomes,
\begin{equation}\label{eq:DC2}
DC=\frac{\left( \bar{f}-F_{\rm baseline}\right)}{<f_{\rm flare}>-F_{\rm baseline}}.
\end{equation}

\begin{figure}[t]
  \centering
  \includegraphics[width=0.5\textwidth]{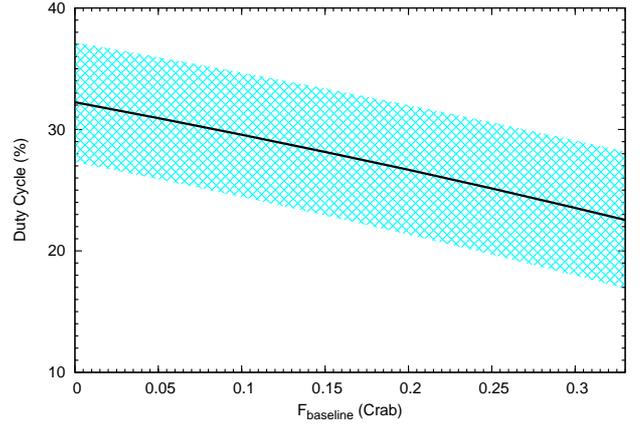}
  \caption{Duty cycle calculated by considering as flaring states all those having a flux above 1 TeV greater than 1 Crab. The shadowed blue area represents the error associated to $DC$, obtained by taking into account the uncertainty on $\bar f$.}
  \label{fig:DC}
 \end{figure}

Equation \ref{eq:DC2} gives the DC as a function of three quantities: 1) the average flux of Mrk 421 ($ \bar{f}$) which has a unique value of 0.85$\pm$0.13 Crab as determined with Milagro observations; 2) the unknown value of the baseline flux ($F_{\rm baseline}$) between 0 and the maximum value of $\sim$ 0.33 Crab and; 3) the average flare flux that depends on the flaring state distribution, $f(x)$ and the  maximum flux ($F_{\rm lim}$) chosen to be 10 Crab. Therefore, we calculated the $DC$, given in Figure \ref{fig:DC}, for values of $F_{\rm baseline}$ from 0 to the upper limit of $\sim$ 0.33 and the uncertainty due to the error associated to $\bar f$. The uncertainty given by the choice of  $F_{\rm lim}$ and $f(x)$ are discussed in Sec. \ref{sec:Fmax} and \ref{sec:f(x)}.

It can be seen from Figure \ref{fig:DC} that the $DC$ ranges from $23^{+5}_{-5}\,\%$ ($F_{\rm baseline}$=0.33 Crab) to $32^{+5}_{-7}\,\%$ ($F_{\rm baseline}$=0 Crab). These values are generally lower but marginally consistent within the error  with the $\sim$40\%\ value obtained by Tluczykont et al. 2007 \cite{2007JPhCS..60..318T}. It is not surprising since, as already explained in Section \ref{sec:intro}, Tluczykont et al. 2007 \cite{2007JPhCS..60..318T} estimated the $DC$ with an observational bias to continue observations of the source in high states, leading to an overestimation of $DC$. 

\subsection[Dependence on $F_{\rm lim}$]{Dependence on $F_{\rm lim}$}\label{sec:Fmax}

The distribution of flux states obtained by Tluczykont et al. 2010 \cite{2010A&A...524A..48T} is based on observations and is not model dependent. The extrapolation of $f(x)$ above 10 Crab is not trivial. For instance,
we do not know if the source can maintain a  flaring state with a flux higher than 10 Crab for a time equal to the duration of the flares in the flux distribution reported in Tluczykont et al. 2010 \cite{2010A&A...524A..48T}. A cut-off in the  distribution is expected at some flux value, just as a result of the limited available energy of the source. Therefore, the extension of $f(x)$ above 10 Crab can not be done much further than 10 Crab. 

 \begin{figure}[t]
  \centering
  \includegraphics[width=0.5\textwidth]{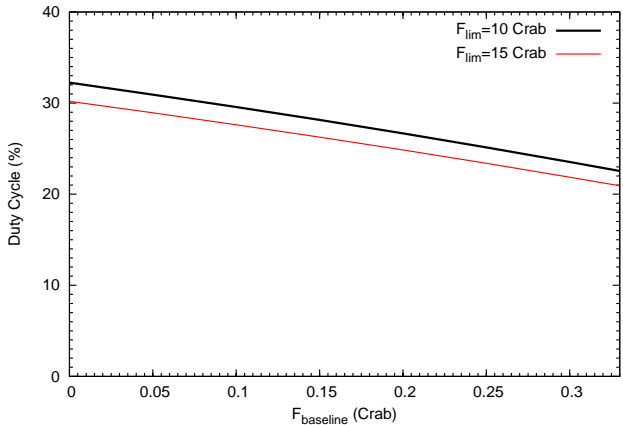}
  \caption{Duty cycle calculated by considering as flaring states all those having a flux above 1 TeV greater than 1 Crab. The black and the red lines correspond to the calculation done by assuming $F_{\rm lim}$=10 Crab and 15 Crab, respectively.}
  \label{fig:DC2}
 \end{figure}

We calculated the $DC$ by considering $F_{\rm lim}$ equal to 15 Crab just by extrapolating the function $f(x)$ up to 15 Crab. The results are shown in Figure \ref{fig:DC2}. It can be seen that, by considering $F_{\rm lim}$=15 Crab the TeV $DC$ goes from 30 \% to 21 \%. These values are between 6\% and 8\% lower than the ones obtained with $F_{\rm lim}$=10 Crab, but are well within the range found including the error on $\bar f$. 


\subsection[Dependence on the distribution of flux states]{Dependence on the distribution of flux states}\label{sec:f(x)}

\begin{figure}[t]
  \centering
  \includegraphics[width=0.5\textwidth]{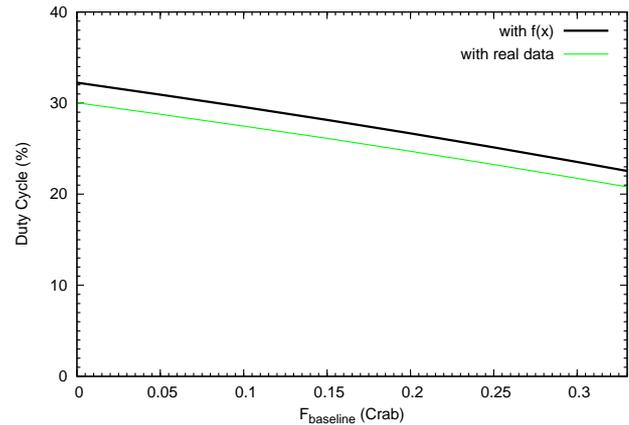}
  \caption{Duty cycle calculated by considering as flaring states all those having a flux above 1 TeV greater than 1 Crab. The black and the green lines correspond to the calculation done by using, for the distribution of flux states, the function $f(x)$ and the discrete set of data (Tluczykont et al. 2010 \cite{2010A&A...524A..48T}), respectively.}
  \label{fig:DC3}
 \end{figure}

The flaring state distribution can be fit by different functions, $f(x)$. We chose the sum of a Gaussian component plus a log-normal function. Tluczykont et al. 2010 \cite{2010A&A...524A..48T} also consider an exponential function above 0.25 Crab. We chose an extreme case to calculate the $DC$. Instead of $f(x)$, we took the actual set of data used to get the distribution of flux states by Tluczykont et al. 2010 \cite{2010A&A...524A..48T}. Then, we obtained a value for  \mbox{$<f_{\rm flare}>$} of 2.83 Crab. The results for the $DC$ are shown in figure \ref{fig:DC3}. It can be seen that, like in the case $F_{\rm lim}$=15 Crab, the TeV $DC$ goes from 30 \% to 21 \%.

\section{Conclusions}
We have presented a new approach to estimate the TeV $DC$ of Mrk 421, based on the continuous monitoring of the source with the Milagro observatory. We have considered the activity of the source above 1 Crab, finding that, depending on the assumed value for the baseline flux of Mrk 421, $DC$ ranges from $23^{+5}_{-5}\,\%$ to $32^{+5}_{-7}\,\%$. These values are  lower but consistent, within the errors, with the value found by Tluczykont et al. 2007 \cite{2007JPhCS..60..318T}. We also tested the robustness of our calculation, finding that the range of values lowers by 6\% - 8\% when a discrete distribution function for the Mrk 421 flux states instead of a continuous function is considered; the same result has been obtained if the distribution is extrapolated up to values of flux greater than the observed 10 Crab. These uncertainties are lower than the one associated to the error on $\bar f$.
 
The value of 1 Crab chosen as flare flux threshold represents an overestimate of the minimum flux required to define a flaring state: in fact, from the distribution of  Tluczykont et al. 2010 \cite{2010A&A...524A..48T} it is clear that above a few tenths of Crab the distribution of flux states presents the typical behaviour of ``high'' states. The estimation of the TeV $DC$ for more realistic assumptions on the threshold flare flux will be presented elsewhere, together with a comparison of the TeV $DC$ with the X-ray $DC$.

\vspace*{0.5cm}
\footnotesize{{\bf Acknowledgment:}{We gratefully acknowledge Scott Delay and Michael Schneider for their dedicated efforts in the construction and maintenance of the Milagro experiment. This work has been supported by the Consejo Nacional de Ciencia y Tecnolog\'ia (under grant Conacyt 105033), Universidad Nacional Aut\'onoma de M\'exico (under grants PAPIIT IN105211 and IN108713) and DGAPA-UNAM.}}


\begin{thebibliography}{}


\bibitem{2007JPhCS..60..318T} M. Tluczykont, M. Shayduk, O. Kalekin and E. Bernardini, Journal of Physics Conference Series 60 (2007) 318.



\bibitem{2008RvMA...20..167H} D. Horns, Reviews in Modern Astronomy 20 (2008) Reviews in Modern Astronomy, ed. S. R{\"o}ser, 167.

\bibitem{2007ApJ...664L..71A} F. Aharonian et al., ApJ 664 (2007) L71.

\bibitem{1995ApJ...440..525V} C. von Montigny  et al., ApJ 440 (1995) 525.


\bibitem{2004ApJ...601..151K} H. Krawczynski et al., ApJ 601 (2004) 151.

\bibitem{2008MNRAS.385..119W} R.M. Wagner,  MNRAS 385 (2008) 119.

\bibitem{2011ICRC....8..147W} S. Wagner, ICRC Proceedings 8 (2011) 147

\bibitem{1991rc3..book.....D} G. de Vaucouleurs, A. de Vaucouleurs, Jr. H.G. Corwin, et al., 1991, Third Reference Catalogue of Bright
398 Galaxies. Volume I: Explanations and references. Volume II: Data for
399 galaxies between 0h and 12h. Volume III: Data for galaxies between
400 12h and 24h.

\bibitem{1992ApJ...401L..61L} Y.C. Lin, et al., ApJ 401 (1992) L61.

\bibitem{1992Natur.358..477P} M. Punch, et al., Nature (1992) 358.


\bibitem{2004ApJ...608..680A} R. Atkins et al., ApJ 608 (2004) 680.

\bibitem{2008ApJ...688.1078A} A. A. Abdo et al., ApJ 688 (2008a) 1078.

\bibitem{2008PhRvL.101v1101A} A. A. Abdo et al., Physical Review Letters 101 (2008b) 221101.


\bibitem{MilagroMrk421} A. A. Abdo et al., ApJ submitted

\bibitem{2010A&A...524A..48T} M. Tluczykont, E. Bernardini, K. Satalecka, et al., A\&A 521 (2010) A48.



\end{thebibliography}
\end{document}